\documentstyle[11pt,aaspp4]{article}
\begin{document}
\title{Identifying the Environment and Redshift of GRB Afterglows from
the Time-Dependence of Their Absorption Spectra}
\author{Rosalba Perna and Abraham Loeb}

\medskip
\affil{Harvard-Smithsonian Center for Astrophysics, 60 Garden Street,
Cambridge, MA 02138}

\begin{abstract}
The discovery of Gamma-Ray Burst (GRB) afterglows revealed a new class of
variable sources at optical and radio wavelengths. At present, the
environment and precise redshift of the detected afterglows are still
unknown. We show that if a GRB source resides in a compact ($\la 100~{\rm
pc}$) gas-rich environment, the afterglow spectrum will show time-dependent
absorption features due to the gradual ionization of the surrounding medium
by the afterglow radiation.  Detection of this time-dependence can be used
to constrain the size and density of the surrounding gaseous system.  For
example, the MgII absorption line detected in GRB970508 should have
weakened considerably during the first month if the absorption occurred in
a gas cloud of size $\la 100~{\rm pc}$ around the source.  The
time-dependent HI or metal absorption features provide a precise
determination of the GRB redshift.

\end{abstract}


%\centerline{submitted to {\it ApJ Letters}, December 1997}

\section{Introduction}

The detection of delayed emission in X-ray (Costa et al. 1997), optical
(van Paradijs et al. 1997, Bond 1997), and radio (Frail et al.  1997)
wavelengths following Gamma-Ray Bursts (GRBs), the so-called
``afterglows'', can be reasonably explained by models in which the bursts
are produced by relativistically expanding fireballs (Paczy\'nski \& Rhoads
1993; Meszaros \& Rees 1997; Vietri 1997a; Waxman 1997a,b; Wijers, Rees, \&
Meszaros 1997; Vietri 1997b; Sari 1997).  On encountering an external
medium, the relativistic shell which emitted the initial GRB decelerates
and converts its bulk kinetic energy to synchrotron radiation, giving rise
to the afterglow. Since the emission occurs on time and length scales which
are much larger than those of the initial explosion, the dynamics of the
expanding shell is self-similar and thus insensitive to the properties of
the energy source (Blandford \& McKee 1976).  The source size inferred from
radio scintillation data of GRB970508 (Frail et al. 1997; Taylor et al.
1997) is consistent with the generic prediction of this model (Goodman
1997; Waxman, Kulkarni, \& Frail 1997).  Moreover, synchrotron emission by
Fermi-accelerated electrons behind the expanding shock can successfully
explain the gross properties of the time-dependent spectra observed in the
afterglows of GRB970228 and GRB970508 for a fireball energy $\sim 10^{52}$
erg (Waxman 1997a,b; Wijers et al. 1997; Sari, Piran, \& Narayan 1997).

In the simplest emission model, the afterglow flux is proportional to the
square root of the ambient medium density, and could therefore span a
dynamic range of more than five decades if the ambient density varies
between its characteristic value in dense molecular clouds ($\sim
10^{4}~{\rm cm^{-3}}$) and in the intergalactic medium ($\sim 10^{-6}~{\rm
cm^{-3}}$).  Even merely galactic environments span a wide range of
possible gas densities, between those of ellipticals and spirals. The
possibility that GRB sources reside in both gas-rich and gas-poor
environments might account for the absence of afterglow emission in some
GRBs (e.g., Groot et al.  1997).  Any direct observational verification of
the postulated connection between the afterglow flux and the GRB
environment would provide a crucial test of the basic fireball model.

Despite the expectation that bright afterglow sources should reside in
gas-rich environments, there is currently no direct measurement of the
density, size, or redshift of the host system for known afterglows.  The
discovery of absorption lines in the optical afterglow of GRB970508
provided the first direct estimate of source distance, constraining its
redshift to the range $0.835\leq z\la 2.3$ (Metzger, et al. 1997;
Djorgovski et al. 1997).  The possible existence of host galaxies has yet
to be conclusively determined from direct imaging of afterglow fields (Sahu
et al. 1997; Fruchter et al.  1997).
%
%The minimum column density expected $\sim 10^{16-17}$ should be sufficient
%to leave HI absorption features by the nearby environment.

If a GRB source resides within the HI disk of a spiral galaxy, its
optical-UV spectrum is likely to show a Lyman-limit trough, a damped
Ly$\alpha$ absorption, and metal absorption lines, all at the source
redshift.  However, the same absorption features could also be produced by
an intervening galaxy along the line-of-sight to the source. The ambiguity
between the GRB host and an interloper could be removed by detecting the
Ly$\alpha$ forest which should end at the true source redshift. However,
for source redshift $z\la 2$, such a spectroscopic detection cannot be made
with ground-based telescopes and requires the use of the Hubble Space
Telescope (hence being feasible only for bright afterglows, due to the
small collecting area).  In this {\it Letter} we argue that if the
absorption occurs in a sufficiently compact region ($\la 100~{\rm pc}$)
around the GRB source itself, then the afterglow radiation will ionize the
region and change the observed absorption features with time. The release
of $\sim 3\times 10^{51}~{\rm ergs}$ in ionizing photons could ionize $\sim
10^{62}$ hydrogen atoms (or $\sim 10^5M_\odot$), and create an ionized
bubble of radius $\sim 100~{\rm pc}~n_1^{-1/3}$ in the surrounding galaxy,
where $n_1$ is the ambient proton density in ${\rm cm^{-3}}$.  Given the
typical HI column densities found in galactic disks (Broeils \& van Woerden
1994), such a bubble could be sufficiently big so as to poke a hole in the
disk and eliminate the absorption features of the host altogether.  Aside
from establishing the GRB redshift, detection of this time-dependent effect
could be used to constrain the gas density and size of the GRB host.  GRB
afterglows are much more effective at ionizing their environments than
supernovae, because their peak UV luminosity is higher by several orders of
magnitude.

The computation of the time-dependence of the absorption spectrum is a
non-trivial problem\footnote{Note that the observed time, which is related
to the width of the outgoing radiation shell, it typically much shorter
than the light crossing time of the absorber.}.  A snapshot of the ionizing
effect that the radiation shell has on the ambient medium is schematically
illustrated in Figure 1. At the late times under consideration, the blast
wave that produced this radiation is already non-relativistic and lags
behind the radiation front. Only the front of the afterglow light pulse
moves into the unperturbed gas.  Each of the subsequent radiation shells
propagates into a medium which had already been partially ionized by prior
shells.  Since each radiation shell has a different spectrum at emission
and a flux that declines with radius as $1/r^2$, the ionization state of
the ambient medium is generally inhomogeneous and time-dependent.  The
eventual absorption spectrum seen by a distant observer, can be calculated
by coupling the radiative transfer equation to the radiative ionization
equation for the medium. In
\S 2, we formulate the set of causal equations that describe the processed
afterglow spectrum, and in \S 3, we present numerical results for the time
evolution of the associated absorption lines.  Our discussion ignores
recombination, since the recombination time ($\sim 10^5~{\rm
yr}~n_1^{-1}$), is typically much longer than the lifetime of the
afterglow.  Our main conclusions are summarized in \S 4.

\section{Ionization Dynamics}

We consider a GRB source which turns on at time $t=0$ and illuminates a
stationary ambient medium of uniform density $n$, with a time dependent
luminosity per unit frequency, $L_\nu(t)$.  Since most of the interstellar
absorption occurs at large radii, we ignore the initial sub-parsec fireball
phase during which the radiation is still coupled to the hydrodynamics.  In
the late phase of interest, the pressure wave has already slowed down to
sub-relativistic speeds, so that it lags behind the radiation field and has
no effect on the absorption spectrum of the outgoing radiation shell. The
ionization state of the gas is then affected only by the luminosity
$L_\nu(t)$ at frequencies higher than the ionization 
threshold $\nu_0$.

The photoionization rate per atom of species $a$ ($=$hydrogen, helium, or
metals) is given by,
\begin{equation} 
-{1\over n_a(r,t)}\frac{dn_a(r,t)}{dt}=\int_{\nu_{0,a}}^\infty d\nu\;
\frac{F_\nu(r,t)}{h\nu}\;\sigma_{{\rm bf},a}(\nu)\;,
\label{eq:dndt}
\end{equation}
where $F_\nu(r,t)=L_\nu(t)/4\pi r^2$ is the flux at a radius $r$ from the
source, and $\sigma_{\rm bf}(\nu)$ is the photoionization (bound-free)
cross-section at a photon frequency $\nu$. All photons are assumed to
propagate radially away from the central source.  Equation~(\ref{eq:dndt})
determines the evolution of the number density of atoms, $n_a(t)$, from
that of the flux, $F_\nu(r,t)$. This flux is absorbed in the medium through
both bound-free and bound-bound transitions.  The radiative transfer
equation in spherical symmetry reads,
\begin{equation}
\frac{1}{r^2}\frac{d[r^2F_\nu(r,t)]}{dr}=- F_\nu(r,t) \sum_{a} n_a(r,t)\; 
\left[\sigma_{{\rm bf},a}(\nu)
+ \sigma_{{\rm bb},a}(\nu)\right]\;,
\label{eq:dJdt}
\end{equation}
where $\sigma_{\rm bb}$ is the cross-section for bound-bound absorption.

The coupled equations (\ref{eq:dndt}) and (\ref{eq:dJdt}) determine the
evolution and spatial distribution of the flux and the ionization fraction,
for a given set of initial and boundary conditions.  We assume for
simplicity that the source is embedded in a spherical gas cloud of uniform
density and radius $R$, namely
\begin{equation}
n_a(t=0,r) = \left\{
\begin{array}{ll}
 n_{a,0} & \hbox{for $r\le R$}\\
  0  & \hbox{otherwise} \\
  \end{array}
 \right.\;,
\label{eq:n0}
\end{equation}
and start the calculation at a radius $r_0\sim 10^{17}~{\rm cm}$
with the boundary condition,
\begin{equation}
F_\nu(r_0,t)=\frac{L_\nu(t)}{4\pi r_0^2}\;.
\label{eq:J0}
\end{equation}
Our results are insensitive to the precise choice of $r_0$, as long as this
radius is much smaller than the scale over which the ambient interstellar
medium is distributed.

Following Waxman (1997a,b), we model the time and frequency dependence of
the afterglow luminosity at emission as
\begin{equation}
L_\nu(t)=L_{\nu_{\rm m}}\left(\frac{\nu}{\nu_{\rm m}(t)}\right)^{-\alpha}\;,
\label{eq:lum}
\end{equation}
where,
\begin{equation}
\nu_m(t)=1.7\times 10^{16}
\left({\xi_e\over 0.2}\right)^2 
\left({\xi_B\over 0.1}\right)^{1/2} 
E_{52}^{1/2}t_{\rm hr}^{-3/2}\;{\rm Hz}\;.
\label{eq:num}
\end{equation}
Here $\xi_B$ and $\xi_e$ are the fractions of the equipartition energy in
magnetic field and accelerated electrons, $E=10^{52} E_{52}~{\rm erg}$ is
the fireball energy, $t_{\rm hr}\equiv (t/{\rm hr})$, and
\begin{equation} 
L_{\nu_{\rm m}}=
%\frac{2}{27\sqrt{2\pi}}\;\frac{\sigma_{\rm T}m_ec}
%{m_p^{1/2}e}\;\sqrt{n_e}\;\xi_{\rm B}\;E\;.
8.65\times 10^{29} \sqrt{n_1}\left({\xi_{\rm B}\over 0.1}\right)E_{52}\;
{\rm {erg\over sec~Hz}},
\label{eq:lumm}
\end{equation}
where $n_1$ is the ambient proton density in units of $1~{\rm cm^{-3}}$.
%, $\sigma_{\rm T}$ is the
%Thomson cross section, $m_e$ and $m_p$ are the electron and proton masses,
%and $e$ is the electron charge. 
In our numerical examples we use the typical parameter values which are
required to fit the existing afterglow data, namely $\xi_e
E_{52}^{1/4}=0.2$, $\xi_B= 0.1$, ${\sqrt{n_1}}E_{52}= 1$, $\alpha= 0.5$ for
$\nu > \nu_m$, and $\alpha=-1/3$ for $\nu < \nu_m$.

Finally, the atomic cross-sections in equation~(\ref{eq:dJdt}) are
calculated as follows.

\noindent
(i) {\it Bound-Bound absorption.} The cross-section for transitions between
two bound states, $i$ and $k$, of an atom has a Lorentzian
shape (Rybicky \& Lightman 1979)
\begin{equation}
\sigma_{\rm bb}^{i,k}=f_{i,k}\;\frac{3\lambda_{k,i}^2}{8\pi}\;
\frac{\Lambda^2_{k,i}(\nu/\nu_{k,i})^2}
{4\pi^2(\nu-\nu_{k,i})^2+\Lambda^2_{k,i}(\nu/\nu_{k,i})^6/4}\;,
\label{eq:sigmaA}
\end{equation}
where $\lambda_{k,i} = c/\nu_{k,i}$ is the transition wavelength, and
$\Lambda_{k,i}$ is the transition probability which is related to the
absorption oscillator strength $f_{k,i}$ through
\[\Lambda_{k,i}=6.670\times 10^{15}~{\rm sec^{-1}}\;g_k\,f_{i,k}/[g_k
(\lambda_{k,i}/{\rm \AA})^2] \;,\] with $g_i$ and $g_k$ being the
statistical weights of the levels $i$ and $k$.  The values of these
parameters for permitted resonance lines of the most important
astrophysical elements are summarized by Verner, Verner \& Ferland (1996).
To minimize the number of free parameters, we assume a cold quiescent
absorber. Thermal or turbulent broadening would change the line profile but
not affect its time dependence, which is dictated by bound-free transitions
[cf. Eq.~(\ref{eq:dndt})].

\noindent 
(ii) {\it Bound-Free absorption.} An analytical expression for the
photoionization cross section exists for single-electron atoms (Hall 1936),
and reads for hydrogen, 
\begin{equation}
\sigma_{bf, \rm H}(\nu)=A_0\left(\frac{\nu_0}{\nu}\right)^4
\frac{\exp{[4-(4\tan^{-1}\epsilon)/\epsilon]}}{1-\exp{(-2\pi/\epsilon)}},
\;\;\; \;\;{\rm for}\;\; \nu\ge\nu_0\;,
\label{eq:sigmaH}
\end{equation}
where $A_0=6.30\times 10^{-18}\;{\rm cm^2}$ and
$\epsilon=\sqrt{\nu/\nu_0-1}$, with $\nu_0=3.29\times 10^{15}~{\rm Hz}$
being the photoionization frequency threshold for an electron in the ground
state. For simplicity, we assume that all neutral hydrogen atoms are in the
ground state; the excitation rate to higher levels does not exceed the
photoionization rate for the characteristic power-law spectrum of GRB
afterglows.

\noindent 
For non-hydrogenic atoms, Verner et al. (1993) obtained a useful analytical
fit to the photoionization cross section $\sigma_{n,l}(\nu)$ for an
electron in a shell with quantum numbers $n,l$. The cross section can be
written as $\sigma_{n,l}(\nu)=\sigma_\star\;F(y)$ for frequencies $\nu\ge
\nu_{n,l}$, where $y\equiv \nu/\nu_\star$.  Here $\nu_{n,l}$ is the
threshold frequency for photoionization of an electron in the shell
${n,l}$; $\sigma_\star=\sigma_\star(n,l,Z,N)$ and
$\nu_\star=\nu_\star(n,l,Z,N)$ are fitting parameters which depend on the
atomic number $Z$ and number of electrons $N$; and 
$F(y)=[(y-1)^2+y_w^2]\,y^{-Q}\left(1+\sqrt{y/y_a}\right)^{-P}$, where
$Q=5.5 + l-0.5P$, and $y_w$, $y_a$, and $P$ are three additional fit
parameters which depend on $n,l,Z$ and $N$. The total cross section to
photoionization is obtained by summing over all shells,
\begin{equation}
\sigma_{\rm bf}(\nu)=\sum_{n,l}\sigma_{n,l}(\nu)\;.
\label{eq:sigmatot}
\end{equation}
We ignore secondary ionization by fast electrons which are created by X-ray
or $\gamma$-ray photoionization.

\section{Numerical Results: Time Dependence of Absorption Lines}

Since hydrogen is by far the most abundant element in galaxies, we examine
first its prominent Ly$\alpha$ absorption signature.  Figure 2 shows the
expected evolution of the afterglow spectrum in a narrow frequency band
around the Ly$\alpha$ resonance for an HI column density of $3\times
10^{20}~{\rm cm^{-2}}$ and two different cases of spherical absorbing
clouds around the source: (i) a compact ($R=1$ pc) and dense ($n_{\rm
H,0}=100$ cm$^{-3}$) cloud (left panels), and (ii) a more extended ($R=100$
pc) and rarefied ($n_{\rm H,0}=1$ cm$^{-3}$) cloud (right panels).  The
first case illustrates the conditions in a molecular cloud, while the
second corresponds to a galactic disk of scale height $\sim 100~{\rm pc}$.
Although the two spectra possess the same absorption profile initially due
to their identical column density, their time evolution is very different.
The smaller absorber is ionized faster, because its atoms are exposed to a
higher ionizing flux and because it requires fewer photons to get ionized
due to its smaller mass.

The photoionization efficiency can be quantified through the change in the
equivalent width ($EW$) of the absorption line, defined as
\begin{equation}
EW=\int d\lambda\;\frac{(F_{\nu,0}-F_{\nu})}{F_{\nu,0}}\;,
\label{ew}
\end{equation}
where $F_{\nu}(\lambda)$ is the {\it observed} spectral flux (at $r\gg R$)
as a function of wavelength across the line and $F_{\nu,0}(\lambda)$ is the
continuum flux that the afterglow would have had if there were no
absorption.  The values of $EW$ for the Ly$\alpha$ line are listed in the
different panels of Figure 2.  For a compact absorbing region such as a
molecular cloud, the $EW$ could decrease by orders of magnitude within
several days after the GRB detection.

Although ground based observations are not capable of detecting Ly$\alpha$
absorption for absorber redshifts $z\la 2$, they can still probe metal
absorption lines as demonstrated in the case of GRB970508 (Metzger et al.
1997; Djorgovski et al. 1997).  Because metal lines are often unsaturated,
it is easier to detect the time evolution in their case than for the damped
Ly$\alpha$ trough.  Figure 3 illustrates the expected evolution of the MgII
absorption line at $\lambda_0=2798\,{\rm
\AA}$, which was detected in the spectrum of GRB970508.
Our calculations assume a solar abundance of MgII (Anders \& Grevesse
1989). Since there are far fewer MgII atoms than HI atoms, it takes fewer
photons and less time from the onset of the afterglow, to ionize them.
Therefore, the evolution of the metal line $EW$ in Figure 3 is more rapid
than that of the Ly$\alpha$ $EW$ in Figure 2.

In Figure 4 we show a broader range of the spectrum, including the UV and
X-ray regimes, where flux is absorbed by photoionization of hydrogen.  For
a $\sim 1$pc absorber, the absorption profile narrows considerably after a
couple of weeks. Consequently, the observed flux from a source at $z_s=1$
in the emission band of 13.6--150eV {\it increases} from $10^{-11}~{\rm
erg~cm^{-2}~s^{-1}}$ after $6~{\rm hr}$ to $4.2\times10^{-11}~{\rm
erg~cm^{-2}~s^{-1}}$ after 2 days, despite the overall fading of the
afterglow emission.

All of the above time dependences occur only for absorbers which are
associated with the GRB source.  More distant absorbers along the
line-of-sight are exposed to an afterglow flux which is much too weak to
affect their ionization state during the brief period of time over which
the radiation lasts.
   
If the GRB environment is metal rich, then it is likely to contain also
dust. The broad-band extinction by dust would suppress the observed
optical-UV flux from the afterglow by a wavelength dependent factor.  The
deviation from the fireball predictions would weaken at late times due to
the heating of dust above its sublimation temperature by the afterglow
radiation (Waxman \& Phinney 1997); this process might explain the
transient suppression of optical relative to X-ray emission during the
first two days of GRB970508.  The absence of an optical afterglow for
GRB970828 might have also been caused by dust extinction (Groot et al.
1997). This interpretation is supported by the low-energy turn-over in the
ASCA X-ray spectrum of GRB970828, which implies a high hydrogen column
density, $\sim 10^{21-22}~{\rm cm^{-2}}$, along the line-of-sight to the
source (Murakami et al 1997).  Figure 4 indicates that if this absorption
originated from a compact molecular cloud around the GRB source, then the
X-ray absorption feature must have evolved on a timescale of days. For a
given inferred column density, the level of evolution becomes weaker with
increasing absorber mass.

\section{Conclusions}

Afterglow models postulate the existence of gas-rich environments around
some GRB sources. We have shown that this conjecture can be tested
empirically by monitoring the afterglow absorption spectra.  If the
surrounding gas is not fully-ionized initially, then the bright afterglow
radiation will modify its ionization state. The medium into which the
afterglow radiation propagates at late times would be more ionized than the
medium into which the early radiation propagated.  Hence, the equivalent
width ($EW$) of the associated absorption lines would decline with time
during the afterglow (see Fig. 2).  The rate of $EW$ variation can be used
to constrain the size of the absorbing region; the more compact the
absorbing region around the GRB source is, the higher is the photon flux to
which it is exposed, and the faster it gets ionized.  When combined with
the value of the column density inferred from the $EW$ measurement itself,
the size estimate can be used to find the density of the ambient
medium\footnote{Note that if the absorber size or density can be measured
independently by another method, then it might be possible to infer the
luminosity distance to the source and by that constrain the values of
cosmological parameters.}.  One could then test statistically the predicted
correlation between the afterglow flux and this density [cf.
Eq.~(\ref{eq:lumm})]. In principle, it might even be possible to infer the
profile of a non-uniform density distribution along the line-of-sight based
on frequent monitoring of the equivalent-width evolution.  Identification
of time-dependence in the afterglow spectrum would isolate the variable
(and hence, associated) absorption lines from other lines which are caused
by chance intersections of unrelated absorbers along the line-of-sight, and
hence provide a precise determination of the GRB redshift.

Our results imply that if the observed metal absorption lines of GRB970508
(Metzger et al. 1997; Djorgovski et al. 1997) occurred in a region of
radius $\la 100~{\rm pc}$ around the GRB source itself, then the equivalent
width of the lines would have declined noticeably over the first month of
the afterglow (Fig. 3).  Similarly, the soft X-ray absorption which was
possibly detected by ASCA in GRB970828 (Murakami et al.  1997) should have
decayed under the same conditions (Fig. 4).  Detection of these effects in
future GRBs would help determine the properties of their gaseous
environment and their redshift. 

Since the peak optical fluxes of afterglows such as GRB970508 are $\ga
10^2$ times brighter than supernovae, they can be used to probe galactic
environments at exceedingly high redshifts. If the GRB source population
traces the star formation history of the Universe, then the dimmest GRB
events known might originate from a redshift as high as $z\sim 6$ (Wijers
et al.  1997).  In hierarchical cold dark matter models for structure
formation, high redshift hosts are expected to be denser [$n
\propto (1+z_f)^3$] and more compact [$R\propto M^{1/3}(1+z_f)^{-1}$, where
$z_f$ is their (higher) formation redshift, and $M$ is their (smaller)
mass] than low redshift galaxies, hence making the effects discussed in
this paper even more pronounced.

\acknowledgements

We thank Ue-Li Pen, John Raymond, Dimitar Sasselov, and Eli Waxman for
useful discussions, and Sarah Jaffe and Max Poletto for help in the
preparation of the figures.  This work was supported in part by a graduate
student fellowship from the university of Salerno, Italy (for RP), and by
the NASA ATP grant NAG5-3085 and the Harvard Milton fund (for AL).

\begin{figure}[t]
\centerline{\epsfysize=5.3in\epsffile{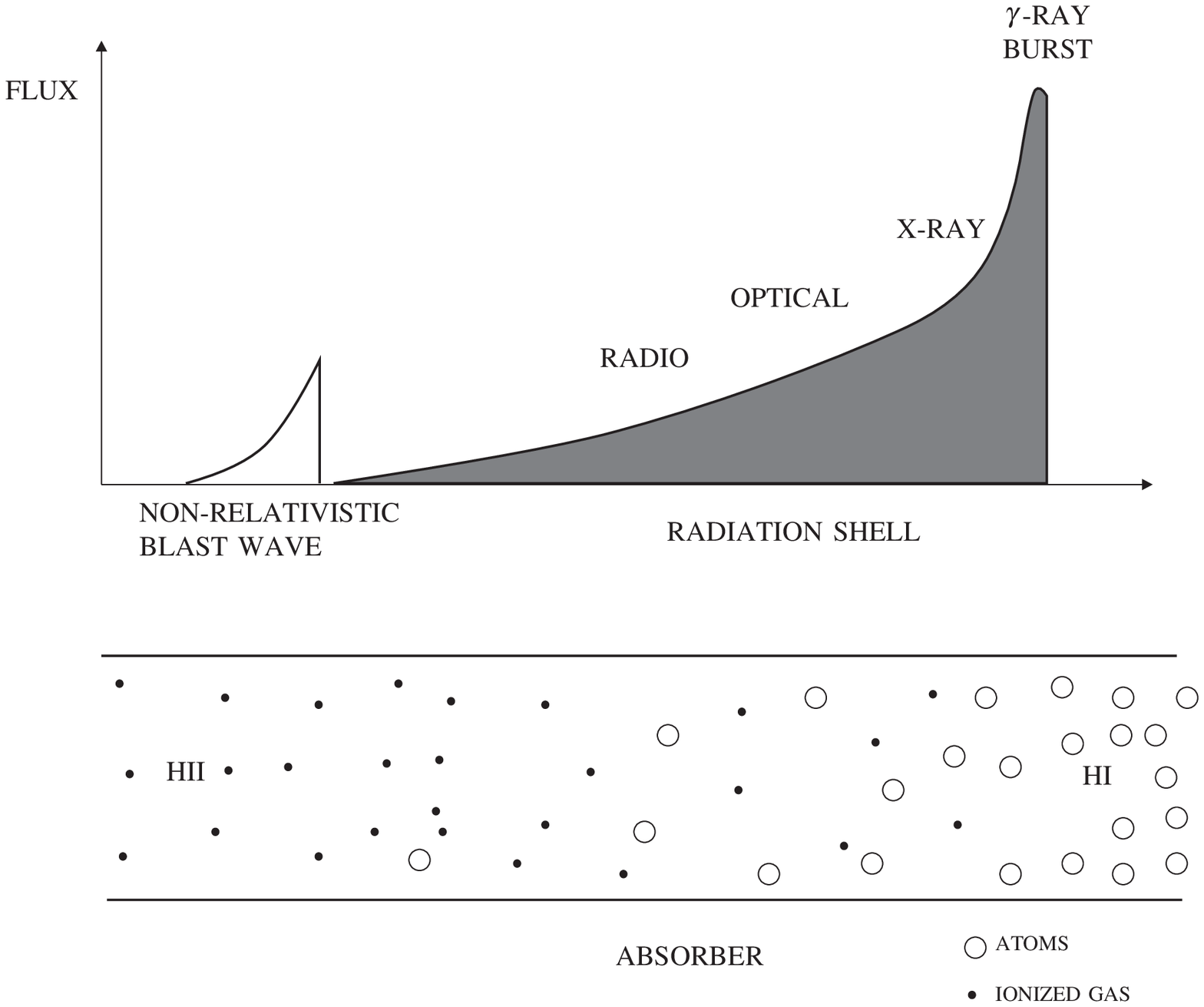}}
\caption{Snapshot of the outward propagating radiation shell 
and its effect on the ambient absorbing medium.  The ambient gas is
gradually ionized by the passing UV radiation. The column of absorbing
atoms seen by the front of the radiation shell is higher than that seen by
the later radiation. As a result, the equivalent width of the absorption
lines declines with time from the point of view of a distant observer.  }
\label{fig:1}
\end{figure}

\begin{figure}[t]
\centerline{\epsfysize=5.7in\epsffile{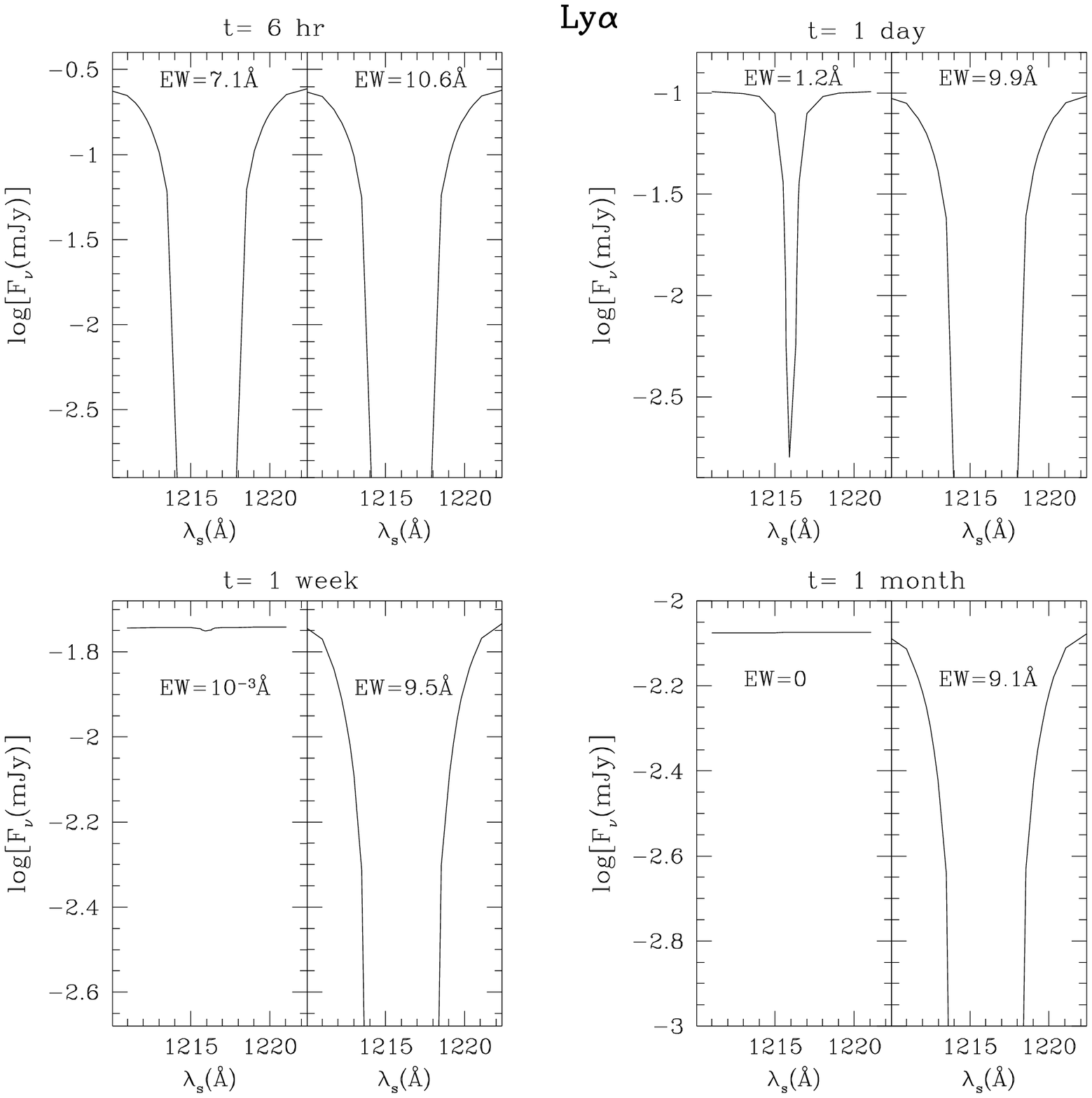}}
\caption{Evolution of the observed Ly$\alpha$ absorption profile for a
spherical gas cloud around a GRB source, with a radius $R$=1 pc and an
initial HI density $n_{\rm H,0}=100~{\rm cm^{-3}}$ (left panels), or with
$R$=100 pc and $n_{\rm H,0}=1~{\rm cm^{-3}}$ (right panels). The equivalent
width ($EW$) of the line is listed in each case. The vertical axis shows
the {\it observed} flux for a source redshift $z_s=1$ (for an $h=0.5$,
$\Omega=1$ cosmology), and the horizontal axis refers to the emission
wavelength at the source, $\lambda_s=\lambda/(1+z_s)$.  }
\label{fig:2}
\end{figure}

\begin{figure}[t]
\centerline{\epsfysize=5.7in\epsffile{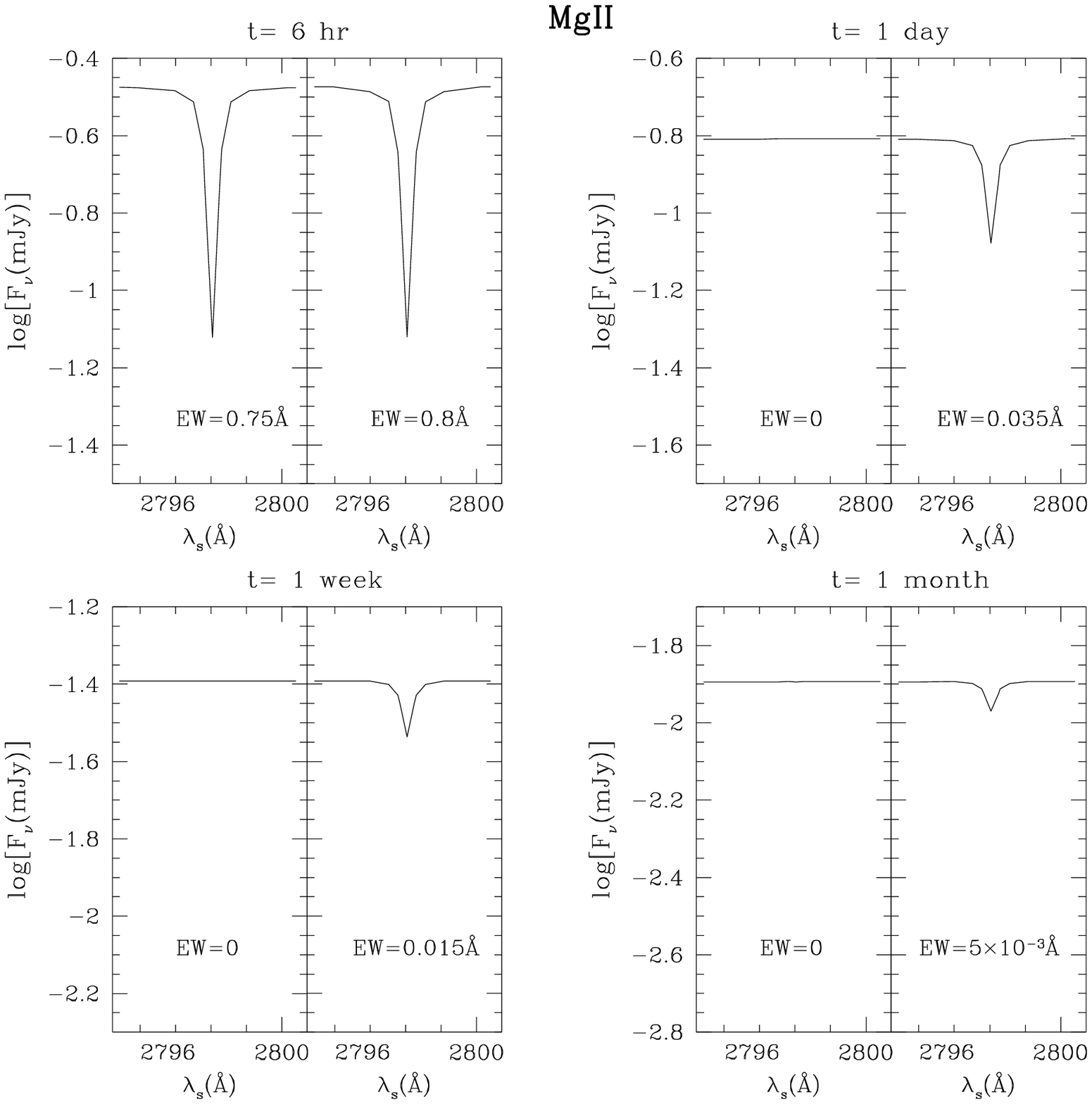}}
\caption{Evolution of the observed MgII absorption line 
%at 2798 $\AA$
for a spherical gas cloud of solar metallicity with $R$=10 pc and an
initial hydrogen density of $n_{\rm H,0}=10~{\rm cm^{-3}}$, or equivalently
a MgII density $n_{\rm Mg,0}=3.8\times 10^{-4} {\rm cm^{-3}}$ (left
panels), as compared to a cloud with $R$=100 pc, $n_{\rm H,0}=1~{\rm
cm^{-3}}$, and $n_{\rm Mg,0}=3.8\times 10^{-5} {\rm cm^{-3}}$ (right
panels). Axis notations are the same as in Fig. 2.}
\label{fig:3}
\end{figure}

\begin{figure}[t]
\centerline{\epsfysize=5.7in\epsffile{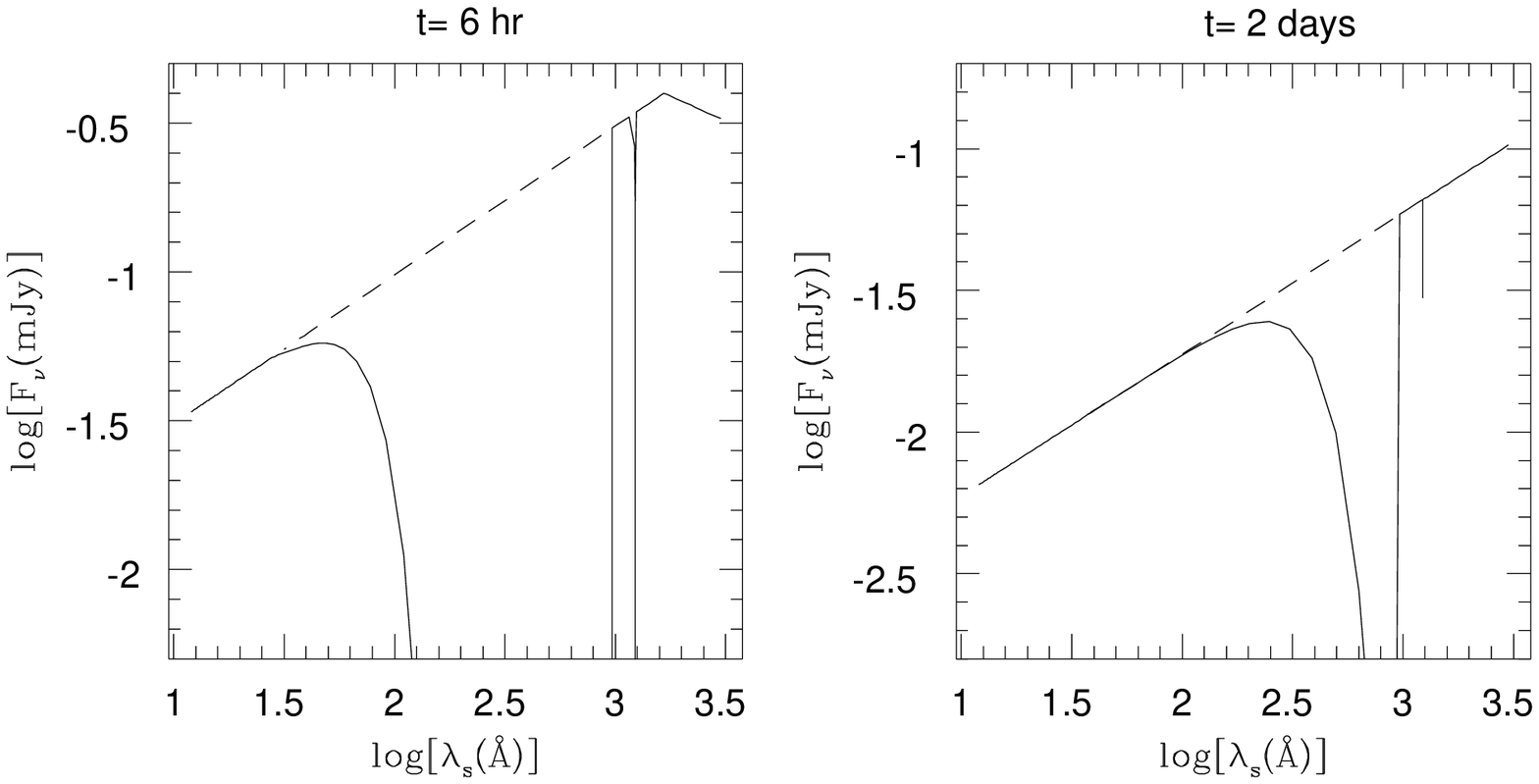}}
\caption{Evolution of the soft X-ray and UV absorption 
for a spherical gas cloud with $R$=1 pc and an initial HI density $n_{\rm
H,0}=100~{\rm cm^{-3}}$.  The dashed line shows the afterglow flux that
would have been observed if there were no absorption. Axis notations are
the same as in Fig. 2.}
\label{fig:4}
\end{figure}

\end{document}